# 3D Wireless: Characterizing Wireless Performance Combining Spatial and Temporal Behaviors

Zhiwei Zhao[†], Wei Dong[†], Jie Yu[†], Tao Gu[‡] and Jiajun Bu[†]
[†]College of Computer Science, Zhejiang University, China
[‡]College of Computer Science, RMIT University, Australia
[†]{zhaozw, dongw, xwolf, bjj}@zju.edu.cn; [‡]tao.gu@rmit.edu.au

*Abstract*—Performance characterization is a fundamental issue in wireless networks for real time routing, wireless network simulation, and etc. There are four basic wireless operations that are required to be modeled, i.e., unicast, anycast, broadcast, and multicast. As observed in many recent works, the temporal and spatial distribution of packet receptions can have significant impact on wireless performance involving multiple links (anycast/broadcast/multicast). However, existing performance models and simulations overlook these two wireless behaviors, leading to biased performance estimation and simulation results.

In this paper, we first explicitly identify the necessary "3-Dimension" information for wireless performance modeling, i.e., packet reception rate (PRR), PRR spatial distribution, and temporal distribution. We then propose a comprehensive modeling approach considering 3-Dimension Wireless information (called 3DW model). Further, we demonstrate the generality and wide applications of 3DW model by two case studies: 3DW-based network simulation and 3DW-based real time routing protocol. Extensive simulation and testbed experiments have been conducted. The results show that 3DW model achieves much more accurate performance estimation for both anycast and broadcast/multicast. 3DW-based simulation can effectively reserve the end-to-end performance metric of the input empirical traces. 3DW-based routing can select more efficient senders, achieving better transmission efficiency.

## I. Introduction

Modeling wireless communication performance is a fundamental problem for various basic functionalities in wireless networks, such as wireless network simulation, routing and flooding protocol designs, etc. For example in network simulation, the goal is to generate packet traces, based on which *the protocol performance* would be similar with that in real world network. In routing protocols [1], [2], [3], the routing decisions are made based on *performance cost* (e.g., expected number of transmissions or delay). Considering that performance metrics such as delay and throughput can be derived using the number of transmissions [4], [5], we consider the number of transmissions as the key performance metric. Therefore, to achieve accurate performance modeling, we focus on estimating the expected number of transmissions (ETX) for various wireless communication modes.

The following are the basic transmission modes for wireless communication: unicast, anycast, multicast/broadcast. Unicast is used for traditional routing protocols such as CTP [6]. Anycast is often used for opportunistic routing protocols [7],

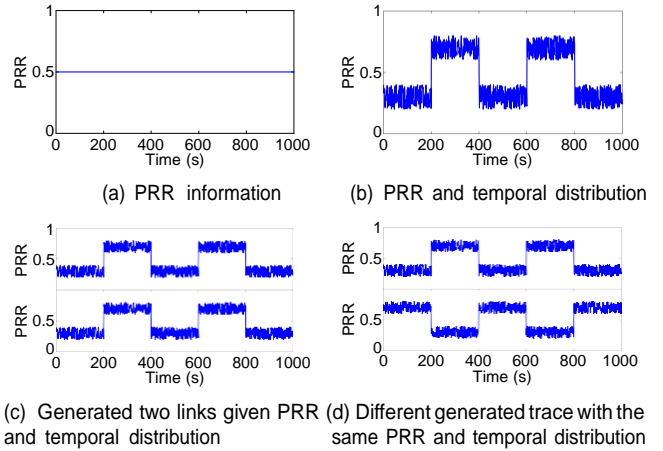

(a) PRR information
(b) PRR and temporal distribution
(c) Generated two links given PRR and temporal distribution
(d) Different generated trace with the same PRR and temporal distribution

Fig. 1. Illustration for the three necessary dimensions for characterizing wireless performance

[8], [9]. Broadcast/multicast are often used for flooding and dissemination protocols [10], [1], [3]. In anycast, the transmission performance is measured based on the transmission number with which at least one receiver receives the packet, denoted as aETX. In broadcast/multicast, the transmission performance is based on the transmission number with which all receivers receive the packet, denoted as bETX.

Modeling unicast performance is simple since it involves only one sender and one receiver. On the other hand, it is much more challenging to model aETX and bETX. Recent studies on wireless link characteristics such as temporal burstiness [11] and spatial link correlation [12] indicate that, packet receptions' spatial and temporal variations can have great impact on protocol performance. Unfortunately, these two kinds of information are overlooked by existing performance modeling approaches [13], [14], [15], [2]. Based on our empirical studies, *we argue that, packet reception rate (PRR), spatial distribution and temporal distribution, are three necessary "dimensions" that should be considered for modeling wireless performance.*

We use an illustrative example, shown in Figure 1, to explain the three characterizing dimensions. If we use only dimension 1 information, i.e., link quality, we can characterize one link with PRR=0.5 as shown in Figure 1(a). Packet reception rates will always be evaluated as 50%. If we use dimension 1 and dimension 2 information, i.e., PRR and temporal distributions, we can characterize one link with PRR=0.5 as shown in

Figure 1(b). We can see that the overall packet reception rate is actually composed of two periodic varying packet reception rates. Such distribution allows us to model fine-grained link performance. So far we have enough information for characterizing one link. Next, we consider two links $l$ and $l^U$. Given the PRR and temporal distribution of these two links, the relationship between $l$ and $l^U$ can be either positively correlated or negatively correlated (Figure 1(c) or 1(d)). The aETX and bETX are largely different for Figure 1(c) or 1(d) [12], [14]. Therefore, the relationship between different links, i.e., the spatial distribution, also has large impact on wireless performance.

Unfortunately, most existing approaches consider only one or two dimensions for performance characterization. These approaches often start from individual link metrics (uETX or packet reception ratio, PRR), and then use the link behaviors of multiple links to further model the transmission performance aETX or bETX. For example, BRE [16] models aETX using PRRs and burstiness, which partially implies the link temporal distributions. It overlooks the spatial correlation among different links. Shuai et al. [2] use a similar approach to model bETX. This approach overlooks the temporal distributions. *Our evaluation results show that*, all the above approaches are not comprehensive enough to characterize wireless behaviors, and hence they cannot achieve accurate model of wireless performance in most cases.

The direct combination of existing characterizations for the three dimensions is challenging. Since the characterizing metrics are individually designed for different dimensions and purposes, it is hard to combine these metrics for deriving the ETX performance. For example, $\beta$ factor [11] is a characterizing metric for temporal burstiness and $\kappa$ factor [12] is a characterizing metric for spatial correlation. These two metrics are designed as normalized values for characterizing temporal and spatial behaviors respectively. However, these metrics are difficult for deriving the wireless performance.

In this paper, combining the identified 3-*D*imension *W*ireless information, we propose a novel characterizing model for wireless performance (called 3DW model). Compared with existing modeling approaches, 3DW efficiently considers PRR, temporal and spatial distributions for uETX/aETX/bETX modeling. We further demonstrate the usefulness of 3DW in two case studies: 3DW-based wireless network simulation and 3DW-based real time routing. For simulation, we devise a multi-level Markov model, which reserves the ETX properties of input traces in the first level and reserves the three-dimension behaviors in the second level. For routing protocols, we incorporate the 3DW model into opportunistic routing [17] and data dissemination [3] protocols, for ETX estimation in the routing decisions.

We conduct extensive simulations and experiments. The results show that 3DW model provides more accurate aETX/bETX estimation than existing models. We also implement 3DW simulation and 3DW routing protocols. Results show that: (1) Compared with existing simulators (TOSSIM [18] and M&M [19]), 3DW simulator can simulate repeatable end-to-end transmission performance, while the other two simulators suffer from large performance variations. (2) 3DW routing protocols outperform their counterparts in terms of transmission count. The reason is that 3DW model provides a more accurate estimation for routing decisions.

The contributions of this paper are summarized as follows:
1) Based on empirical and analytical results, we identify the three-dimension necessary information for wireless performance modeling.
2) We propose a performance modeling approach (3DW) which efficiently considers all the 3D essential wireless information. Evaluation results show that 3DW achieves more accurate aETX/bETX estimation than existing modeling approaches.
3) We incorporate 3DW model into wireless network simulation and real time routing protocols. Evaluation results show that 3DW indeed enhances the performance of both simulation and routing protocols.

The rest of the paper is organized as follows: Section II presents the discussion of impacting factors on wireless performance. Section III introduces the state-of-the-art modeling approaches and their limitations. Section IV presents the 3DW modeling approach. Section V and VI present the two case studies on simulation and real time routing protocols. Section VII evaluates the 3DW model as well as the 3DW simulation and 3DW routing. Section VIII concludes the work and points future directions.

## II. CHARACTERIZING DIMENSIONS

In this section, we analyze the impacting factors of wireless communication performance and conclude the necessary three-dimension information required for characterizing wireless unicast/anycast/broadcast/multicast performance.

### A. Packet reception rate

Link quality is perhaps the most widely identified impacting factor for wireless performance. Packet reception ratio (PRR) is a typical characterizing metric. It is a link-wise long term property, indicating the probability that a packet can be successfully received.

Figure 2(a) shows an empirical trace of a wireless link. If we use a PRR value to characterize this link, we can generate the expected PRR as shown in Figure 2(b). Obviously, the PRR variations are not characterized. A step further, we would like to see whether the wireless performance can be characterized. The expected number of transmissions for successfully delivering one packet (ETX) is used as the performance metric. To obtain the ground truth ETX, we account the number of losses before a packet reception, $n_{loss}$. Then the number of transmissions for a packet reception is $n_{tx} = n_{loss} + 1$. Then we average the number of transmissions to obtain the ETX value. The experiment is repeated 100 times. Figure 2(c) shows the CDF of ETX values for both empirical and the generated traces. We can see that, the ETX values are largely different, i.e., the generated traces can hardly characterize the ETX performance. The reason is that PRR

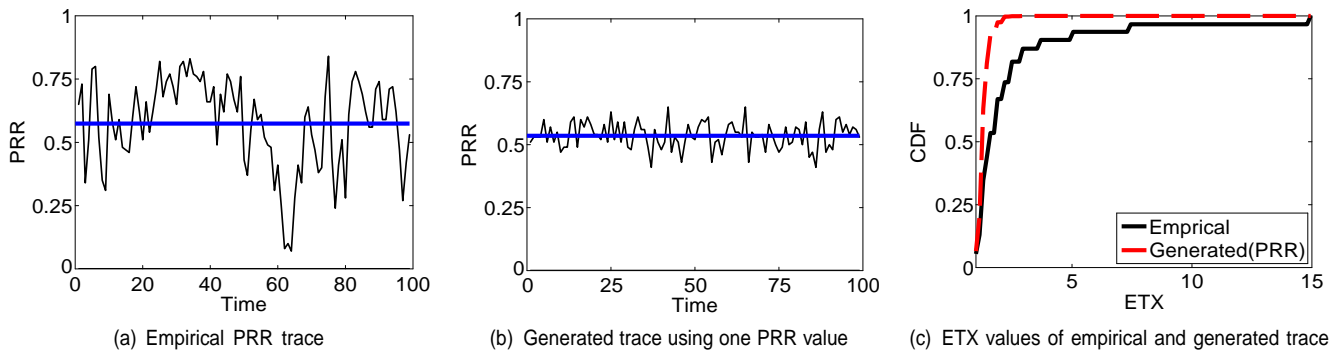

Fig. 2. Empirical studies on packet reception rates (PRR).

(a) Empirical PRR trace
(b) Generated trace using one PRR value
(c) ETX values of empirical and generated trace

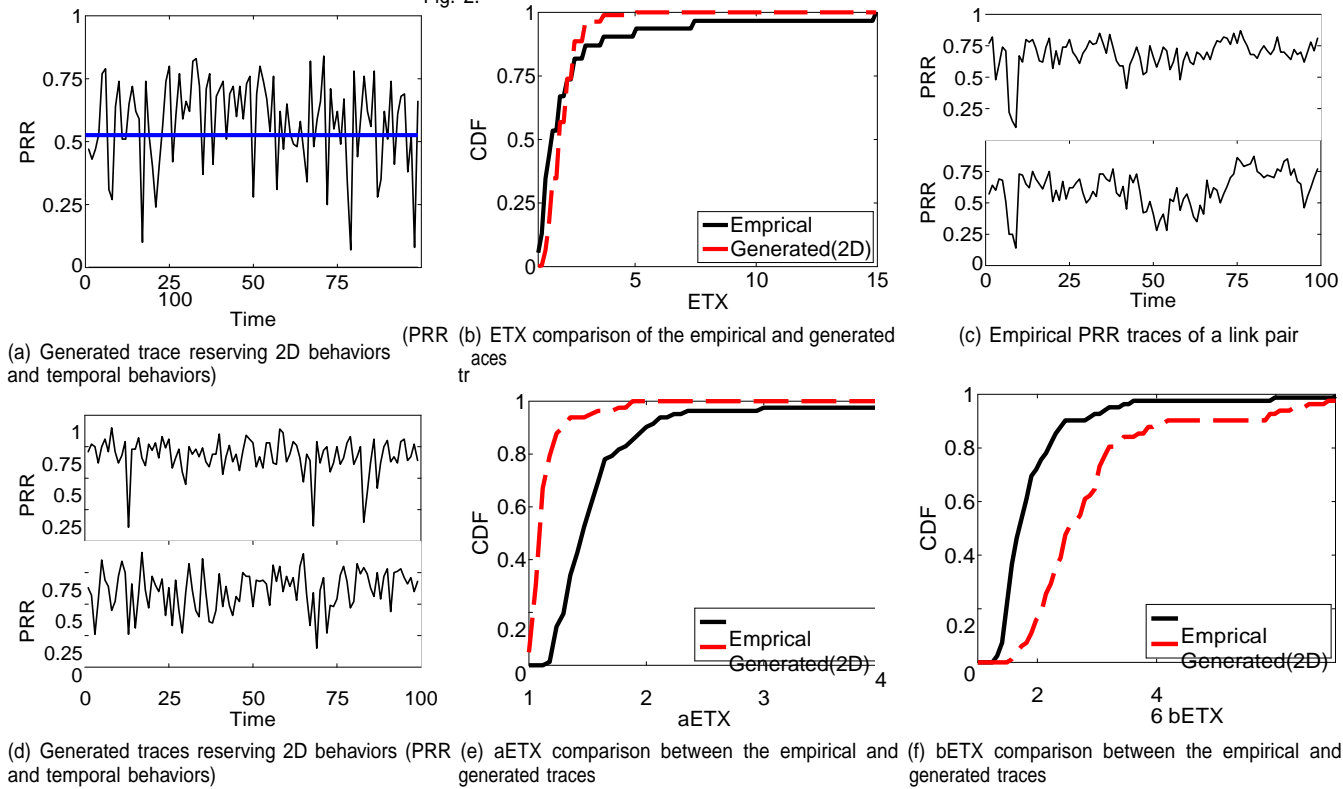

(a) Generated trace reserving 2D behaviors (PRR and temporal behaviors)
(b) ETX comparison of the empirical and generated traces
(c) Empirical PRR traces of a link pair
(d) Generated traces reserving 2D behaviors (PRR and temporal behaviors)
(e) aETX comparison between the empirical and generated traces
(f) bETX comparison between the empirical and generated traces

Fig. 3. Empirical studies on temporal behaviors.

captures long term link behaviors, i.e., the overall ETX of the two traces are similar. However, due to the PRR variations, the fine-grained short term performance cannot be captured, resulting in the inaccurate performance characterization. From this experiment, we find that PRR alone cannot accurately capture the ETX performance.

Some works have tried to exploit PHY layer information [20], [21] or machine learning algorithms [22] for achieving real time PRR estimation. However, as we will introduce in the next subsection, these approaches are trying to characterize the temporal distributions of packet reception behaviors, which is denoted as the second dimension (temporal distribution) for characterizing wireless performance.

### B. Temporal distribution

It has been observed by many recent works that both packet reception and lost have clear temporal behaviors [23], [11], [16]. These works try to characterize the temporal distribution using various metrics such as $\mu$, $\beta$, etc.

We calculate the PRR and temporal PRR distributions of empirical traces (Figure 2(a)). Then we use the Markove model approach in [19] to generate packet traces for both links, reserving both PRR and the temporal packet distributions. Figure 3(a) shows generated PRR trace, reserving the long-term PRR and temporal distributions. Figure 3(b) shows the ETX comparison between the empirical trace and the generated trace. We can see that ETX is much more accurately characterized. Reserving the two dimension properties (i.e., link quality and temporal distribution) seems enough for characterizing single link performance.

Next, we study whether the wireless performance involving multiple links (anycast/broadcast/multicast) can be characterized. Figure 3(c) shows the empirical PRR traces of two links. We can see that these two links have high positive

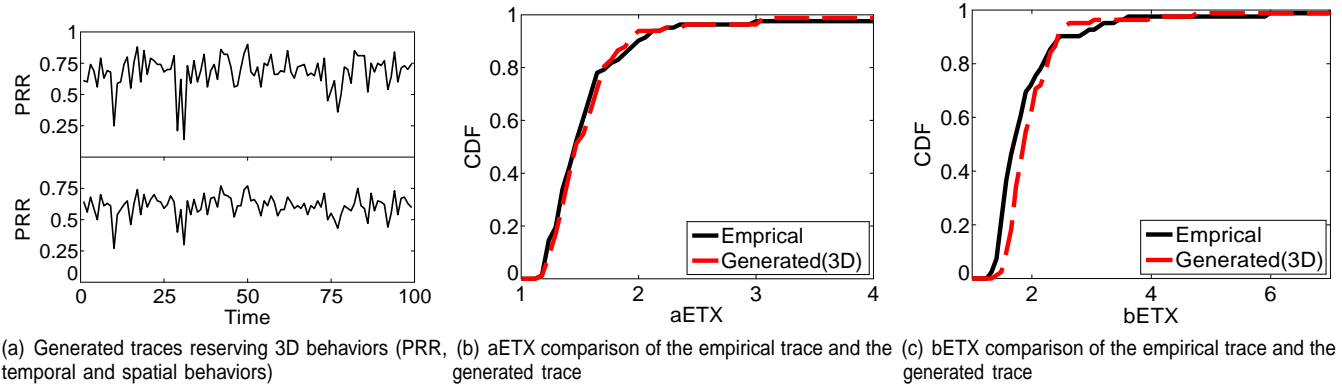

(a) Generated traces reserving 3D behaviors (PRR, temporal and spatial behaviors)
(b) aETX comparison of the empirical trace and the generated trace
(c) bETX comparison of the empirical trace and the generated trace

Fig. 4. Empirical studies on spatial behaviors.

correlation. Figure 3(d) shows the generated PRR and ETX traces reserving PRR and temporal distributions. Different with the empirical traces, there is no clear correlation between the two generated PRR traces. Then we would like to see whether the end to end performance is characterized. We study the performance in terms of aETX and bETX. Using the traces for the two links, we can easily count the number of transmissions for delivering one packet to at least one receiver and the number of transmissions for delivering one packet to both receivers. Figure 3(e) shows the aETX values for empirical traces and generated traces. We can see that aETX is not accurately characterized. Similarly, as shown in Figure 3(f), bETX is neither accurately characterized. The reason is that, anycast and broadcast can be greatly affected by spatial distributions. If two links receptions have positive correlation, aETX tends to be large [14] and bETX tends to be small [12] for the same generated packet traces on both links.

### C. Spatial distribution

The spatial distribution of PRRs essentially reflects the relationship among different links, which has been observed by many existing works [24], [12]. Now we manually generate PRR traces, reserving PRR, temporal and spatial distributions. Figure 4(a) shows the generated traces. We can see though the individual PRR variations are similar with 3(d), the relationship between two links are different. Figure 4(b) shows the CDF of link correlation values of empirical trace pair and generated trace pair. We can see that the CDF of link correlation is very similar. Next we would like to see the end to end performance. Figure 4(c) shows the CDF comparisons of aETX and bETX. We can see that both aETX and bETX are much more accurately characterized.

### D. Short Summary

In high level, PRR characterizes the long term property of a link. The temporal distribution characterizes the temporal properties of PRR variations. These two dimensions can accurately characterize single link performance. The spatial distribution characterizes the relationship between different links. With the three dimensions, we can probabilistically assure the uniqueness of a generated trace; Each link's PRR variations and its relationship with other links can be determined, which essentially determines the anycast/broadcast/multicast performance.

Therefore, only when the three dimension information is given, we are able to *accurately* infer the wireless behaviors as well as wireless performance in terms of uETX, aETX and bETX.

Although some works have considered one or two of the above three dimension information, to the best of our knowledge, this is the first analytical work that concludes all necessary information for characterizing wireless behavior and performance.

### III. STATE-OF-THE-ART MODELING APPROACHES

In this section, we introduce the related works, with focus on the state-of-the-art performance models. Considering modeling of unicast is straightforward, we mainly introduce the modeling approaches for aETX/bETX. We would like to discuss whether existing models consider the identified three dimension information for performance modeling, and further compare the modeling accuracies of our proposed 3DW model with these works.

#### A. Anycast

Given a sender $s$ and a set of receivers $S_R$, the expected number of transmissions for $s$ to cover at least one node in $S_R$ is calculated as:

$$aETX = \frac{1}{P^s_{S_R}} \quad (1)$$

where $P^s_{S_R}$ is the probability that at least one node in $S_R$ receives the packet. Existing modeling approaches differ from each other mainly in the calculation of $P^s_{S_R}$.

**Modeling using PRR information.** In [15], the authors consider PRR information for modeling ETX for anycast, i.e., aETX.

$$P^s_{S_R} = 1 - \prod_{i \in S_R}(1 - p_{Si}) \quad (2)$$

**Modeling using PRR and link correlation** In [14], the authors consider both PRR and link correlation for modeling aETX. The key difference is the calculation of $P^s_{S_R}$, the

probability that at least one node in $S_R$ receives the packet. In [14], $P^s_{S_R}$ is calculated as:

$$P^s_{S_R} = \sum_{k=1}^{n}(-1)^{k-1}Pr(f^k) \quad (3)$$

where $f^k \subset F = f_1,...,f_n$ is any candidate forwarder set with size $k$, and $Pr(f^k)$ is the probability that the $k$ candidate forwarders successfully receive a packet.

*B. Broadcast & Multicast*

Given a sender $s$ and a set of receivers $S_R$, the bETX is calculated as:

$$bETX = \prod_{k=1}^{+\infty} kP(X=k) \quad (4)$$

where $P(X = k)$ denotes the probability that all nodes in $S_R$ are covered after $k$ transmissions. Existing modeling approaches differ from each other mainly in the calculation of $P(X = k)$.

**Modeling using PRR only.** In [13], the authors consider PRR information for modeling bETX. The probability that all $M$ receivers receive the packet after k transmissions, $P(X = k)$ is given by:

$$P(X=k) = \prod_{i=1}^{M}(1-p_i^k) - \prod_{i=1}^{M}(1-p_i^{k-1}) \quad (5)$$

Then bETX is calculated as:

$$bETX = \sum_{k=1}^{+\infty} k(\prod_{i=1}^{M}(1-p_i^k) - \prod_{i=1}^{M}(1-p_i^{k-1})) \\
= \sum_{i_1,i_2,...,i_M} \frac{(-1)^{i_1+i_2+...+i_{M}-1}}{1-p_i^{i_1}p_2^{i_2}...p_M^{i_M}} \quad (6)$$

From the above equations, we can see that the input is $M$ separate PRRs ($p_n^{in}$). Both the temporal distribution and spatial distribution are not considered.

**Modeling using PRR and link correlation.** In [2], a bETX modeling approach is proposed. The authors assume that in one transmission round from one sender to multiple receivers, the receivers of better PRR links receive the packet earlier than other links. Based on this assumption, the bETX is modeled as:

$$bETX = \frac{1}{p(e_1)} + \frac{Pr(\overline{e_2}|p(e_1))}{p(e_2)} + ... + \frac{Pr(\overline{e_M}|\bigcap_{i=1}^{M-1}p(e_i))}{p(e_M)} \quad (7)$$

where $p(e_1)$ is the PRR of link 1, i.e., the packet reception rate of receiver 1; $Pr(\overline{e_2}|p(e_1))$ stands for the probability that receiver 2 loses the packet when receiver 1 receives the packet.

We can see that, the above model considers both PRR and link correlation, which is a metric implying the spatial distribution of packet reception. The temporal distribution is yet not considered.

**Summary** We can see that the above modeling approaches consider either one dimension or two dimensions for performance modeling. More specifically, we observe that some works [14], [2] have already considered the spatial link behaviors (link correlation) but failed to consider all three dimensions. However, modeling wireless performance exploiting all three dimensions is a non-trivial task. The simple combination of existing characterizations is difficult. Since the characterizing metrics are separately designed, these metrics are difficult to be combined. For example, $\beta$ factor is a characterizing metric for temporal burstiness and $\kappa$ factor [12] is a characterizing metric for spatial correlation. These two metrics themselves are designed complex, such that directly exploiting them for performance modeling is almost infeasible. Different from these approaches, the proposed 3DW modeling does not rely on existing metrics and instead takes a matrix as input information, which efficiently convey the three dimension wireless behaviors.

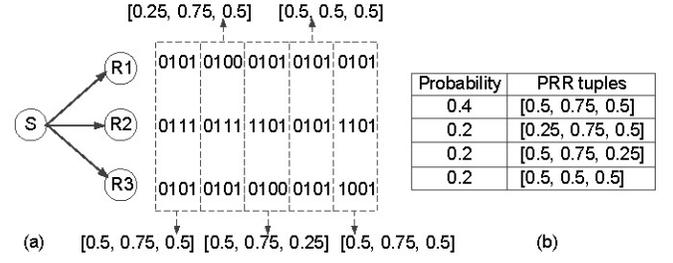

Fig. 5. Illustration of the 3DW model input acquisition.

We will compare the accuracies of the above modeling approaches in Section VII.

IV. THE 3DW MODELING APPROACH

In this section, we present the 3D Wireless (3DW) Performance modeling approach, which efficiently considers three dimension information and achieves more accurate wireless performance modeling.

*A. Model input*

As we have discussed in Section III, different from existing modeling approaches, the input of 3DW is not explicit characterizing metrics such as link correlation or burstiness. Instead, we extract a PRR distribution from the packet reception traces on multiple links, which stores the temporal and spatial distributions of PRR values. With the distribution, the three kinds of information are jointly considered in the input probability set.

Given packet reception traces on different links, we first slice time into many short periods, and obtain a series of PRR values for each link. After that, we combine PRR values at the same period in a PRR tuple, and account the overall probability of each different PRR tuples. After that, we obtain a table storing PRR tuples and its corresponding probabilities. One different PRR tuple represents one different spatial distribution case. The probabilities for PRR tuples represent the temporal distributions and variations.

Figure 5(a) shows an illustrating example, where $S$ is a sender, $R1,R2,R3$ are three receivers. The "0" and "1" represent packet losses and receptions. Taking the packet reception traces as input, we first slice the traces into several windows (each window contains 4 packets). In each window, we can obtain a PRR tuple, e.g., the PRR tuple of the first window is [0.5,0.75,0.5], indicating that PRRs on the three links are 0.5, 0.75 and 0.5 within the same window. After that, we can obtain the probabilities of all different PRR tuples as shown in Figure 5(b). This table is the input for 3DW modeling. We can see that PRR values, spatial distributions and temporal distributions are all covered by the input table.

### B. aETX

Similar with existing works, the aETX is calculated as:

$$aETX = \frac{1}{p_{S_R}^s} \quad (8)$$

where $p_{S_R}^s$ is the probability that at least one node in $S_R$ receives the packet. Since we separately account different states, we can directly obtain $p_{S_R}^s$ as follows:

$$\begin{aligned} p_{S_R}^s &= 1 - p_{0*} \\ &= 1 - \sum_{\forall t_i \in T} p(t_i) \prod_{i=1}^{n} p_{t_i}(0*) \end{aligned} \quad (9)$$

where $p_{0*}$ denotes the probability all receivers lose the packet ("0" stands for a packet loss), $p(t_i)$ denotes the probability of PRR tuple $t_i$, and $p_{t_i}(0*)$ denotes the probability that all receivers lose the packet given PRRs in tuple $t_i$.

### C. bETX

Recall that bETX is the expected number of transmissions for a sender to cover all its receivers. The receivers are not restricted to receive the packet at the same time.

For simplicity, we start from the case of three receivers R1, R2 and R3. Basically, bETX can be calculated as:

$$bETX = \sum_{k=1}^{+\infty} k P(X = k) \quad (10)$$

where $P(X = k)$ is the probability that k transmissions cover all three receivers. It can be calculated as

$$P(X = k) = P(X > k - 1) - P(X > k) \quad (11)$$

where $P(X > k)$ is the probability that after k transmissions, at least one receiver have not received the packet.

The calculation of $P(X > k)$ turns out to be an inclusion-exclusion problem as shown in Figure 6. Note that P(R1=0) denotes the probability that *after k transmissions*, R1 has not received the packet, P(R1=0&R2=0) denotes the probability that after k transmissions, both R1 and R2 have not received the packet, and P(R1=0&R2=0&R3=0) denotes the probability that after k transmissions, R1, R2 and R3 have not received the packet. With the above information,

$$\begin{aligned} P(X > k) &= P(R1=0) + P(R2=0) + P(R3=0) - \\ & P(R1=0\&R2=0) - P(R1=0\&R3=0) - P(R2=0\&R3=0) + \\ & P(R1=0\&R2=0\&R3=0) \\ &= P_{n_0=1} - P_{n_0=2} + P_{n_0=3} \end{aligned} \quad (12)$$

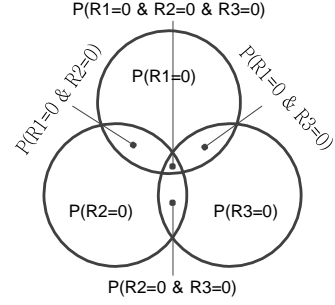

Fig. 6. Calculation of bETX: The case of three receivers.

where $P_{n_0=1}$ denotes the probability that $n_0(=1)$ receivers lose the packet k times. With the input, we get:

$$\begin{aligned} P_{n_0=1} &= (p_{000}+p_{001}+p_{010}+p_{011})^k + \\ & (p_{000}+p_{001}+p_{100}+p_{101})^k + \\ & (p_{000}+p_{010}+p_{100}+p_{110})^k \\ P_{n_0=2} &= (p_{000}+p_{001})^k + \\ & (p_{000}+p_{010})^k + \\ & (p_{000}+p_{100})^k \\ P_{n_0=3} &= (p_{000})^k \end{aligned} \quad (13)$$

Combining Eq. (10)-(13), we can obtain the bETX to cover the three nodes.

**n-receivers case**. Now we move to calculate the bETX to cover n receivers. In high level, it can be similarly calculated as Eq. (10). The key is to calculate $P(X > k)$, the probability that not all n receivers are covered after k transmissions. We use a n-bits bitmap to denote the case of packet reception distribution. For example, a bitmap of "0101" denotes the case in which the first and third receivers lose the packet and the second and the fourth receivers receive the packet. Then $P(X > k)$ is given as:

$$\begin{aligned} P(X > k) &= \sum_{m=1}^{n} (-1)^{m-1} P_{n_0=m} \\ &= \sum_{m=1}^{n} (-1)^{m-1} \sum_{\forall S_m} e_{S_m}^k \end{aligned} \quad (14)$$

where $P_{n_0=m}$ is the probability that m receivers are not covered by k transmissions, $S_m$ is set of bitmaps with m "0"s and $e_{S_m}$ is the probability with m uncovered receivers. $e_{S_m}$ is calculated as:

$$e_{S_m} = \sum_{\forall b \in S_m} p_b = \sum_{\forall b \in S_m} \sum_{\forall t_i \in T} p_{t_i}(b) \quad (15)$$

where $b$ is a bitmap with $m$ 0s and $p_{t_i}(b)$ denotes the probability of the bitmap $b$, given the PRR tuple of $t_i$.

Combining Eq. (10), (11) and (14), the bETX to cover n

receivers is given by:

$$bETX = \sum_{k=1}^{+\infty} kP(X=k)$$

$$= \sum_{k=1}^{+\infty} k(P(X>k-1) - P(X>k))$$

$$= \sum_{k=1}^{+\infty} k \sum_{m=1}^{n} (-1)^{m-1} (\sum_{\forall S_m} (e_{S_m}^k - e_{S_m}^{k-1})) \quad (16)$$

$$= \sum_{m=1}^{n} (-1)^{m-1} \sum_{\forall S_m} \sum_{k=1}^{+\infty} k(e_{S_m}^{k-1} - e_{S_m}^k)$$

$$= \sum_{m=1}^{n} (-1)^{m-1} \sum_{\forall S_m} \frac{1}{1-e_{S_m}}$$

**Short discussion.** We can see that the three characterizing dimensions are considered in the proposed 3DW model. Intuitively, for approaches that do not consider spatial/temporal distribution, it implicitly assumes random spatial/temporal distribution. Therefore, it can achieve accurate model only when the ground truth are of spatial/temporal distributions. We will compare the modeling accuracies of 3DW with other approaches in Section VII.

## V. Exploiting 3DW for Wireless Simulation

Different from existing simulation approaches, which try to repeat the PRR properties or temporal properties, We argue that the key to wireless simulation is to the ability to replay the wireless performance.

To achieve this goal, we propose a novel 3DW-based multi-level Markov model for wireless network simulation. The framework is depicted in Figure 7. In high-level, we use a Markov model to capture the network performance variations. We first identify the performance states of the measured traces, in terms of aETX and bETX. Each state is a tuple of (aETX,bETX), which is obtained from measured packet traces using 3DW model. The high-level Markov model is driven by a (aETX,bETX) transition matrix.

For each specific high-level (aETX,bETX) state, we devise a low-level Markov model to generate the packet receptions and losses at each different receiver. The low-level Markov model is driven by a PRR tuple transition matrix.

### A. The Double-level Markov model

The first level captures the long term aETX and bETX variations. Each state is a tuple of (aETX,bETX). The state transitions represent the long term performance variations. The second level captures the PRRs' spatial and temporal distributions. Each state is a PRR tuple, indicating the PRR values at different receivers. We can see that in the second level model, a state essentially represents the PRRs' spatial distributions. The transition matrix represents the temporal variations. Finally, packet receptions at multiple receivers are concurrently generated, according to the PRR tuple.

It is worth mentioning that wireless communication is inherently based on broadcast and packet receptions/losses happen at the same time. Hence, compared with existing

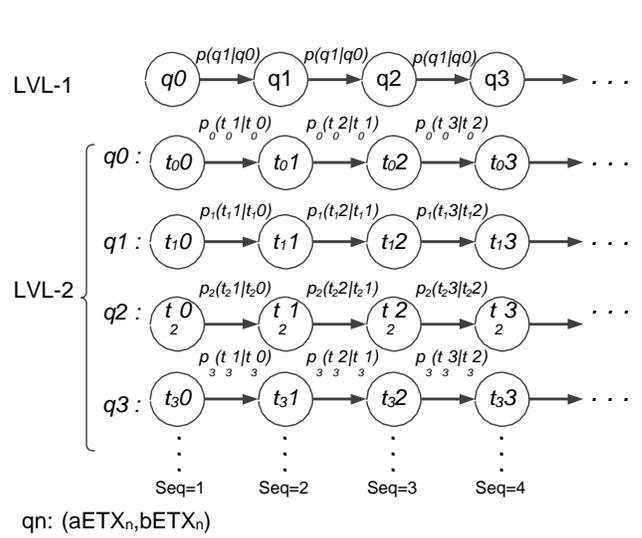

qn: (aETX$_n$, bETX$_n$)
t$_m$n: a n-elements PRR tuple indicating PRR distributions at different receivers, given the high level state of qm

Fig. 7. The double level Markov model. The first level captures the (aETX,bETX) transitions. The second level captures the PRR tuple transitions, where

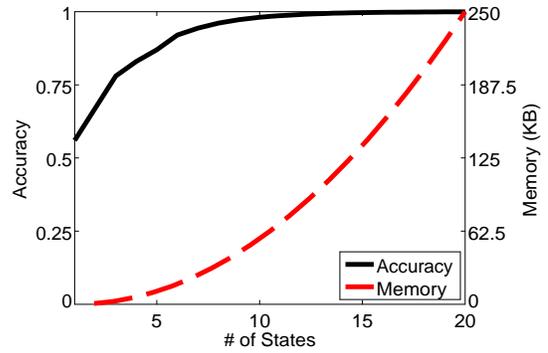

Fig. 8. Impact of states number on modeling accuracy and memory overhead.

"link-wise" approaches, which separately simulate each link's packet receptions, 3DW simulation is more reasonable in the simulation manner.

**State clustering.** To reduce the memory overhead, we can reduce the number of different states in both Level-1 and Level-2 Markov models. For example in level-1 Markov model, if we divide the range of aETX/bETX values into 10 sections, there will be 10x10 different (aETX,bETX) states. The transitional matrix size will be 100x100. To reduce the number of states, we use the k-means clustering to cluster all (aETX,bETX) values into k clusters. The cluster centers can then be used as the input states for Level-1 model.

In our experiments, each sender has three receivers on average, and the link quality range contains five values, 0, 0.25, 0.5, 0.75, 1. Figure 8 shows the model estimation accuracies and memory overhead with varying k values. The tradeoff is clear: with a small k, the estimation accuracies decrease and memory overhead decreases. With a large k, the estimation accuracies increases yet the memory overhead increases. In our experiments, we select a k=5, because it achieves relative high accuracy (above 0.8) while incurring smaller memory overhead (15.6KB). Similarly, the Level-2 model states can also be optimized by clustering the traces into k states.

*B. Comparison with M&M model*

M&M [19] is the state-of-the-art work of wireless simulation. The key insight of M&M model is that it considers both long term and short term PRR variations. A multi-level Markov model is proposed, where the first level captures the long term PRR variation and the second level captures the short term PRR variation.

Compared with M&M, 3DW has the following main differences:

- M&M aims to simulate the long term and short term *packet reception rates*, i.e., uses the PRR transitions to generate bit sequence on each link. The underlying claim is that M&M tries to reserve the PRR property of the packet traces. Differently, in 3DW simulation, we consider the end-to-end metric aETX and bETX. The transition of tuples of (aETX,bETX) is used to generate the bit sequences on multiple links. As a result, 3DW simulation can reserve the property of aETX and bETX, which is more important for network protocol simulation.
- M&M considers only link-wise packet reception variations, which overlooks the spatial distributions of packet receptions. As discussed in Section II, spatial distribution can greatly affect the wireless performance in terms of broadcast/multicast and anycast. Hence when simulating broadcast/multicast/anycast operations, M&M may suffer from large performance variance. Differently, 3DW efficiently considers all three dimension information needed that have impact on wireless performance.

*C. Comparison with TOSSIM*

TOSSIM [18] is a typical wireless simulator. It does not directly generate PRR values. Instead, it generates noise values at different receivers and generate packet receptions according to the SNR (Signal to Noise Ratio) model. The noise trace is generated according to historical noise readings, using the Closest Pattern Matching model.

Compared with TOSSIM:

- TOSSIM greatly relies on the accuracy of SNR model, which can be largely different on different network node radios [25]. Conversely, 3DW simulation extracts the network behaviors directly from the packet-level traces, which is a more practical reflection of the real world wireless performance.
- 3DW simulation generates receptions at multiple receivers at the same time, which can effectively maintain the spatial and temporal distributions of packet receptions.
- 3DW provides the wireless performance repeatability while TOSSIM provides only repeatability of the link-wise receptions.

## VI. EXPLOITING 3DW FOR REAL TIME ROUTING

The main benefit brought by 3DW for real time routing is that it provides an accurate metric for evaluating the aETX/bETX, which fundamentally supports the routing decisions in various network protocols. More specifically, the estimation of aETX can be used for selecting efficient forwarders in opportunistic routing. The estimation of bETX can be used for selecting efficient senders in broadcast protocols (flooding[26]/dissemination[10]/etc.). It is also worth noting that once the ETX values are obtained, the delay and throughput can be further calculated [4]. The incorporation of 3DW is straightforward: Using 3DW model for calculating the aETX/bETX, supporting the routing decisions. However, considering the limited resource of wireless nodes, the main challenge is to reduce the memory and computation overhead.

**aETX.** For a sender with n receivers, the input would be $m^{2n}$ probabilities if we divide the range of PRR into m sections. Apparently, the overhead is considerably large, especially for dense wireless networks.

From the modeling approach of aETX, we can see that only $p_{0*}$ is accounted, where $p_{0*}$ is the probability that all receivers lose the packet. Therefore, when used for opportunistic routing, only $p_{0*}$ is required. Compared with existing approaches that uses link quality [15] or link quality/correlation [14], the approach greatly reduces the memory overhead.

**bETX.** From Eq. (16), we can see that in the calculation of bETX, all bitmap cases with $m$ 0s are taken for calculation. Recall that a bitmap indicates the packet receptions at different receivers. $m$ denotes the number of receivers that lose the packet. To reduce the complexity, we devise a simple approximation method, deliberately reducing the number of different bitmap cases used for calculation. Then the approximated bETX, $b\hat{ETX}$ is calculated as:

$$b\hat{ETX} = \sum_{m=1}^{n} (-1)^{m-1} \sum_{i=0}^{c} \frac{1}{1 - e_{S_m^U}} \quad (17)$$

where $c$ is the manually set number of bitmap cases, and $S_m^U$ is a random bitmap set with $m$ receivers with packet losses. When $c$ is set large, the bETX calculation would be more accurate but the computation overhead will be large. When $c$ is set small, the calculation would be less accurate and the computation overhead will be small.

## VII. EVALUATION

In this section, we evaluate the proposed 3DW modeling approach as well as the simulation and real-time case studies.

*A. 3DW Modeling*

We use one sender and two receivers to study the model accuracies of both aETX and bETX. We manually tune the PRR, temporal and spatial distributions of PRR, and compare the estimation accuracies. 3DW models of aETX and bETX are separately evaluated, in terms of accuracy and computation overhead.

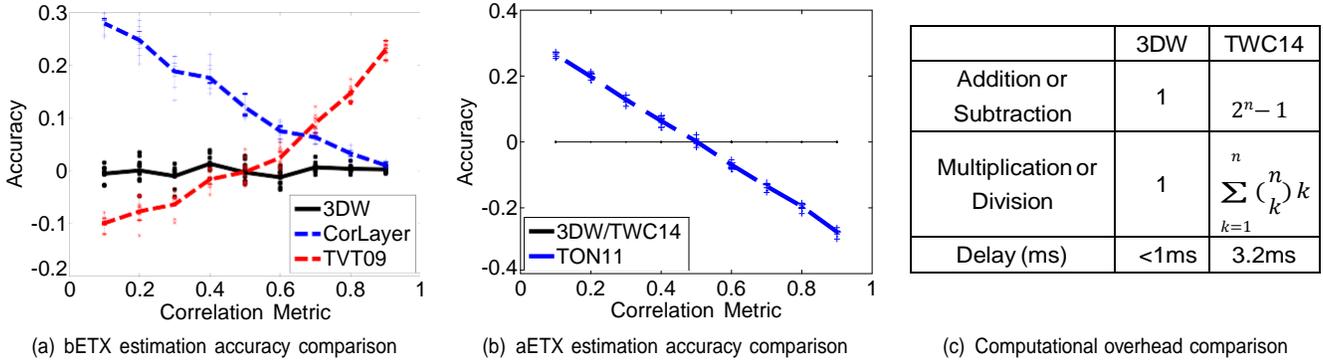

(a) bETX estimation accuracy comparison  (b) aETX estimation accuracy comparison  (c) Computational overhead comparison

Fig. 9. Evaluation results on the 3DW modeling approach.

**Baseline approaches.** For aETX, we compare 3DW with two existing works [15], [14]. One work considers only link quality (Eq. (2)) and the other work considers both link quality and link correlation (Eq. (3)). For bETX, we also compare 3DW with two existing works [13], [2]. One work [13] considers only link quality (Eq. (6)) and the other work considers both link quality and link correlation (Eq. (7)).

Figure 9(a) compares the bETX estimation accuracies of 3DW and other approaches. We can see that, (1) TVT09 [13] is accurate only when link correlation is around 0.5, i.e., the spatial distribution is random. The reason is that it does not consider the spatial distributions, and implicitly assumes that the PRRs are randomly distribution at different receivers. (2) CorLayer [2] is accurate when link correlation is strong and inaccurate when link correlation is weak. The reason is that it is based on the assumption that receivers of better quality links receive the packets earlier than other receivers. When link correlation is 1, it can be treated as all receivers receive the packet at the same time, which minimizes the impact of the assumption.

Figure 9(b) compares the aETX estimation accuracies. We can see that (1) 3DW is more accurate than the approach in [15]. The reason is that 3DW identifies and considers more information, thus achieving more accurate result. (2) 3DW and TWC14 have the same accuracy. The reason is that TWC14's modeling (Eq.(3)) essentially takes $2^n$ link correlation values, which implicitly takes the spatial distribution as well as its temporal distributions. Therefore, the the estimation results are as accurate as 3DW.

Since the accuracies of 3DW and TWC14 are the same, we next compare the estimation overhead of 3DW and TWC14. Figure 9(c) compares the computation overhead of 3DW and TWC14 [14]. We can see that TWC14 require much more operations than 3DW. The reason is that the probability of all zeros can be directly obtained by $p_{00}$ (Eq.(9)). When implemented in the MSP430 platform, 3DW's calculation delay is much more less than that of TWC14.

*B. 3D Wireless simulation*

We use our 8x10 TelosB nodes testbed to collect packet traces. Each node periodically broadcast packets and records the packet receptions from neighboring nodes. The packet receptions on each link are sent to the PC via USB cables. Using the collected packet traces at all links, we can then simulate the network using 3DW-simulation. For comparison, we also use the measured traces to drive the M&M model. Similarly, we measure the noise traces on our testbed, and fed it into the TOSSIM simulation.

We run the popular data dissemination protocol, Deluge [10], on both real sensor testbed as well as in the three simulators. Then we compare the end to end performance in terms of the total number of transmissions (NTX) during the dissemination process. We run the protocols 1000 times.

Figure 10(a) compares the relative-error CDF plots of different simulation approaches. We can see that, most of the NTX simulation errors of 3DW are smaller than 0.03. Conversely, Only about 50% cases in M&M simulates the NTX with relative errors smaller than 0.03. The performance of TOSSIM is even worse.

The reason that 3DM outperform the other two simulators is two-fold: (1) 3DW deliberately reserve the ETX properties in its Level-1 Markov model. The state transitions characterizes the ETX variations, which further supports 3DW-simulation to generate similar ETX performance. (2) 3DW is more comprehensive in that it considers the three necessary characterizing dimensions. Such that the ETX performance properties can be maintained.

*C. Real time routing performance*

When used for real time routing, as discussed in Section VI, an optimized lightweight version of 3DW performance modeling is incorporated. We focus on opportunistic routing (ORW [17]) and dissemination (CoCo [3]). For ORW, we change the forwarder set selection module, and select the forwarder set with the smallest aETX, estimated by 3DW. For CoCo, we change the sender selection module, and select the sender with the smallest bETX.

Similarly, we exploit the baseline approaches ([14] for aETX and [2] for bETX) into ORW and CoCo as well. Then we compare the number of transmissions of 3DW-based protocols and the baseline protocols.

Figure 10(b) shows the performance of opportunistic routing exploiting different performance models. Figure 10(c) shows the performance of dissemination exploiting different performance models. We can see that for both protocols, 3DW-based approaches achieves the least number of transmissions.

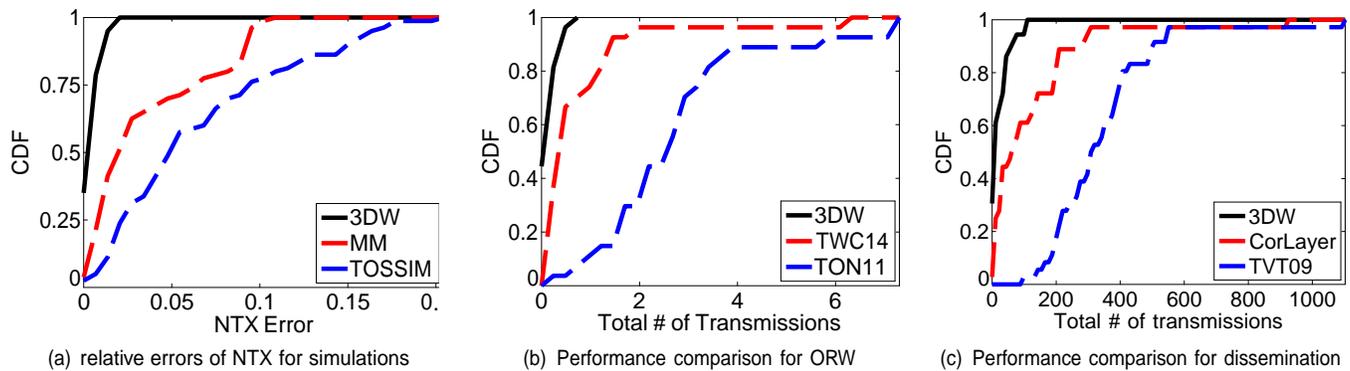

Fig. 10. Evaluation results on the two case studies: 3DW-simulation and 3DW-based real time routing.

The reason is that 3DW model provides the most accurate aETX/bETX estimation, thus achieving the most accurate routing decisions in both opportunistic routing and dissemination.

## VIII. CONCLUSION

In this paper, we study the problem of characterizing wireless performance. we explicitly identify the necessary "3-Dimension" information for wireless performance modeling, i.e., packet reception rate, spatial distribution and temporal distribution. Then we propose a comprehensive modeling approach considering 3D Wireless information (called 3DW model). After that, we demonstrate the generality and fundamental applications of 3DW model with two case studies: 3DW-based network simulation and 3DW-based real time routing protocol. 3DW simulation can reserve the performance properties while 3DW-real time routing can efficiently select more efficient forwarders during the routing process.

We conduct extensive simulation and testbed experiments. Results show that 3DW model achieves much more accurate performance estimation for both anycast and broadcast/multicast. 3DW-based simulation can effectively reserve the end-to-end performance metric of the input empirical traces. 3DW-based routing can select more efficient senders, achieving better transmission efficiency.